\documentclass[prl,twocolumn,superscriptaddress,floatfix]{revtex4}
\usepackage{epsfig,amsmath,amssymb,latexsym}

\newcommand{\beq}{\begin{eqnarray}}
\newcommand{\eeq}{\end{eqnarray}}
\newcommand{\siesta}{\textsc{Siesta}}
\newcommand{\transiesta}{\textsc{Tran\-siesta}}

\usepackage{subfigure}

\begin{document}

\title{Atomic carbon chains as spin-transmitters: an \textit{Ab initio} transport study}
\preprint{1}
\author{Joachim A. F\"{u}rst}
\email[Corresponding author: ]{joachim.fuerst@nanotech.dtu.dk}
\affiliation{DTU Nanotech -- Department of Micro- and
Nanotechnology, NanoDTU, Technical University of Denmark, DK-2800 Kongens Lyngby, Denmark}
\author{Mads Brandbyge}
\affiliation{DTU Nanotech -- Department of Micro- and
Nanotechnology, NanoDTU, Technical University of Denmark, DK-2800 Kongens Lyngby, Denmark}
\author{Antti-Pekka Jauho}
\affiliation{DTU Nanotech -- Department of Micro- and
Nanotechnology, NanoDTU, Technical University of Denmark, DK-2800
Kongens Lyngby, Denmark} \affiliation{Aalto University, Department of Applied Physics,  P.O.\ Box 11100, FI-00076 AALTO, Finland}
\date{\today}

\begin{abstract}
An atomic carbon chain joining two graphene flakes was recently
realized in a ground-breaking experiment by Jin  {\it et al.}, Phys.
Rev. Lett. {\bf 102}, 205501 (2009). We present {\it ab initio}
results for the electron transport properties of such chains and
demonstrate complete spin-polarization of the transmission in large
energy ranges. The effect is due to the spin-polarized zig-zag edge
terminating each graphene flake causing a spin-splitting of the
graphene $\pi_z$ bands, and the chain states. Transmission occurs
when the graphene $\pi$-states resonate with similar states in the
strongly hybridized edges and chain. This effect should in general
hold for any $\pi$-conjugated molecules bridging the zig-zag edges
of graphene electrodes. The polarization of the transmission can be
controlled by chemically or mechanically
modifying the molecule, or by applying an electrical gate.
\end{abstract}
\maketitle
\par{\it Introduction.}
Carbon based materials have assumed a central role in the
development of nanoelectronics \cite{Avouris2007}. The spectacular
electronic properties of Carbon nanotubes and 2D graphene sheets 
are exploited in many device proposals possibly paving the way to an
era of carbon-based electronics. Further, the magnetic properties of
these structures make them promising candidates in spintronics and
numerous suggestions have been made for carbon based spin-filtering
systems
\cite{Ferrer2009,Tombros2007,Furst2009,Vanevic2009,Bhamon2009,Yazyev2009,Fujita1996,Nakada1996,Wang2008b,
Yazyev2008,Munoz-Rojas2009,Furst2008,Koleini2007}.

The fabrication of a field effect transistor by carving a
constriction in graphene \cite{Geim2007} has triggered a massive
interest in such geometries. Graphene nanoribbons forming the
constrictions have been studied intensively. Already some time ago
Nakada {\it et al.} \cite{Nakada1996} showed that these ribbons have
peculiar edge states which depend on the edge geometry as well as
the ribbon width. Zig-zag edges are particularly intriguing as they
support localized spin-polarized states. For example, Mu\~noz-Rojas
{\it et al.} predict a giant magnetoresistance in a device based on
zig-zag ribbons connected to metal electrodes
\cite{Munoz-Rojas2009}. The thinnest possible constriction, the
monoatomic linear chain, was already several years ago proposed
theoretically as an ideal component in molecular devices
\cite{Lang1998,Larade2001,Tongay2004}. Until recently these chains
have been studied only in very few experiments due to the daunting
challenges in their fabrication
\cite{Lagow1995,Zhao2003,yuzvinsky2006}.

Recently Jin {\it et al.} \cite{Jin2009} realized quite stable freestanding carbon
chains connected to two graphene flakes. By employing energetic electron irradiation
inside a transmission electron microscope two holes were created in a large graphene
sample. The resulting ribbon separating the two holes was carefully thinned by lower
energy electrons until only the monoatomic chain remained. This structure is thus both
the realization of a pure carbon constriction of the thinnest possible kind, as well as
an atomic chain contacted by carbon leads instead of the usual case of metallic
contacts. Chuvilin {\it et al.} have reported similar results \cite{Chuvilin2009}. Jin
{\it et al.} also performed density functional theory (DFT) studies showing that the
zig-zag edge is the energetically most favorable structure for connecting the chain, in
agreement with the experimental observations. Recently Chen {\it et al.}
\cite{Chen2009a} reported calculations of the electronic transmission through this
system based on a tight-binding $\pi$-orbital model; the dominant features were
asymmetric resonant peaks  close to the Fermi level.
\begin{figure}[tbh]
\begin{center}
\includegraphics[width=0.8 \columnwidth,angle=0,viewport= 0 200 700 460,clip]{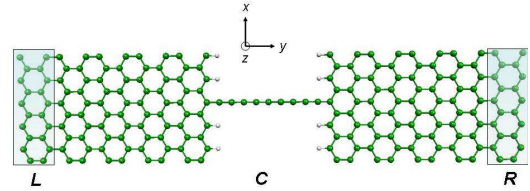}
\caption{(Color online) The C-9 system with 9 carbon atoms in the chain.
The shaded areas marked \textit{L} and \textit{R} are the electrodes,
and \textit{C} is the central region.
Periodic boundary conditions are imposed in the transverse, \textit{x}, direction.}
\label{chain_system}
\end{center}
\end{figure}

In this Letter, we describe the spin-dependent electronic transport
properties of carbon chains connecting two graphene sheets at the
zig-zag edges. From $ab$ $initio$ calculations we find a strong
spin-polarization of the transmission caused by the intrinsic
properties of the zig-zag edge: it spin-polarizes and introduces a
spin splitting of the graphene $\pi$ bands as well as the hybridized
edge-chain states. The peaks in the transmission arising as graphene
states are in resonance with the strongly hybridized edge and chain
states are thus separated in energy resulting in the polarization of
the current. We argue that this is a generic feature of this class
of systems making them strong candidates for future spintronic
devices.

\par{\it Method and Systems.}--We use $ab$ $initio$ pseudopotential spin-polarized DFT as implemented in the \siesta ~code \cite{Soler2002} to obtain
the electronic structure and relaxed atomic positions. Our spin transport calculations
are based on the nonequilibrium Green's function method \cite{Haug2008} as implemented in
the \transiesta\cite{Brandbyge2002} code, extended to spin-polarized systems
\footnote{We employ GGA PBE for exchange-correlation \cite{Perdew1996}, mesh cutoff of
200 Ry, and Monkhorst-Pack k-sampling of (4,1,10). The transmission is averaged over 100
$k_x$-points. The basis set size is single-$\zeta$ in transport calculations, and we checked
that larger double-$\zeta$ does not have qualitative influence on the transmission.}.

The model system is inspired by that of Ref. \cite{Jin2009}
and is shown in Fig. \ref{chain_system} in the case of $N =$ 9
carbon atoms in the chain, denoted C-9. The system is divided into
left ($L$) and right ($R$) electrodes, and a central region
\textit{C} \cite{Brandbyge2002}. The electrodes are semi-infinite
and we employ periodic boundary conditions in the transverse,
\textit{x}, direction. Most importantly, the two graphene sheets are
terminated by zig-zag edges. Initially, we consider them to be
H-passivated.
\begin{figure}[t!]
\begin{center}
\includegraphics[width=0.98 \columnwidth,angle=0,viewport=0 3 730 570,clip]{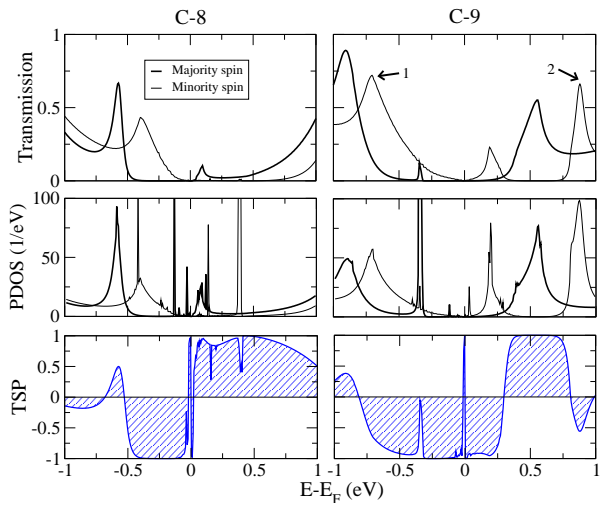}
\caption{(Color online) (Top): The electron transmission of the C-8
and C-9 systems. (Middle): PDOS of chain atom $\pi_z$-orbitals.
(Bottom): The spin-polarization of the transmission shown above
given in terms of TSP.} \label{gmr}
\end{center}
\end{figure}
We use a fixed C-C distance of 1.3 \AA~ in the chain. We find no
qualitative differences in the transmission from using fully relaxed
odd $N$ structures \footnote{We have performed relaxation with a
double-$\zeta$ basis set on the C-9 system using the conjugant
gradient method and a force tolerance of 0.01 eV/\AA. Only the chain
atoms and the two closest rows of atoms on each side were allowed to
move. We find bonding lengths in agreement within 0.05 \AA~ of Ref.
\cite{Jin2009}.}. For even $N$ a careful relaxation migh result in a dimerisation in the chain as found by Khoo \textit{et al.}\cite{Khoo2008}. To
test whether this affects  the transmission, we have computed the
transmission with fixed single/triple bond lengths of 1.26/1.36 \AA~
in the chain of the C-8 system, as discussed below.
\par{\it Results.}\label{sec_res} The spin-resolved transmission is shown in
Fig. \ref{gmr}(top), for two systems with different chain lengths. Both cases show large
peaks away from the Fermi level $E_F$, for both spin channels and smaller peaks closer
to $E_F$. All peaks have asymmetric line-shapes in agreement with Chen {\it et al.}
\cite{Chen2009a}. At $E_F$ the transmission is zero. Outside the energy window
$|$E$|<1$ eV shown in Fig. \ref{gmr}, large peaks are found repeatedly at higher and
lower energies, but with much less pronounced spin-splitting. We focus here on the range
$|$E$|<1$ eV around the intrinsic Fermi-level, which can be modified by  chemical doping
\cite{Novoselov2007a}, or by a gate voltage \cite{Geim2007}. In this range we obtain
very similar transmissions for C-7 and C-9, and for C-8 and C-10, respectively. Thus, a
clear odd/even effect is present as in other carbon chain systems
\cite{Lang1998,Larade2001}. The introduction of a rigid dimerisation for the C-8 system causes minor relative 
shifts in the resonance peaks but the qualitative picture is unaltered (these results
are available in the Supplementary material S1). We stress that a
detailed study of odd/even effects is not our primary concern here, and that the
comparisons discussed here only serve to demonstrate that the spin-polarization
persists in both cases.

The total transmission is spin-polarized as the peaks of each spin channel
are shifted relative to each other. This is illustrated in Fig.
\ref{gmr} (bottom), where we plot the spin-polarization of the
transmission (TSP) given by \beq \hbox{TSP} = \frac{T_{\rm
Maj}-T_{\rm Min}}{T_{\rm Maj}+T_{\rm Min}}. \eeq Significant energy
windows above and below $E_F$ display complete polarization,
$|$TSP$|$ $\approx 1$. The polarization changes sign around $E_F$
being predominantly minority (majority) spin below (above) $E_F$. The majority (minority) spin is defined as having the highest (lowest) electron occupation. 

We next examine the features in the transmission in more detail. An
infinite monatomic carbon chain has two spin-degenerate transmitting
channels; one channel consists of the $\pi_z$-orbitals and the other
one of the $\pi_{x}$-orbitals not participating in the $sp$
hybridization in the chain. Since transport takes place via the
$\pi_z$-orbitals in the graphene sheets and since they are
orthogonal to the $\pi_x$-orbitals, transmission will occur only
through the $\pi_z$-orbitals of the system. This explains why the
peak height is close to unity for each spin channel, as seen in the transmission spectra
of the total system in Fig. \ref{gmr} (top).

\begin{figure}[t!]
\begin{center}
\includegraphics[width=0.8 \columnwidth,angle=0,viewport=150 0 520 520, clip]{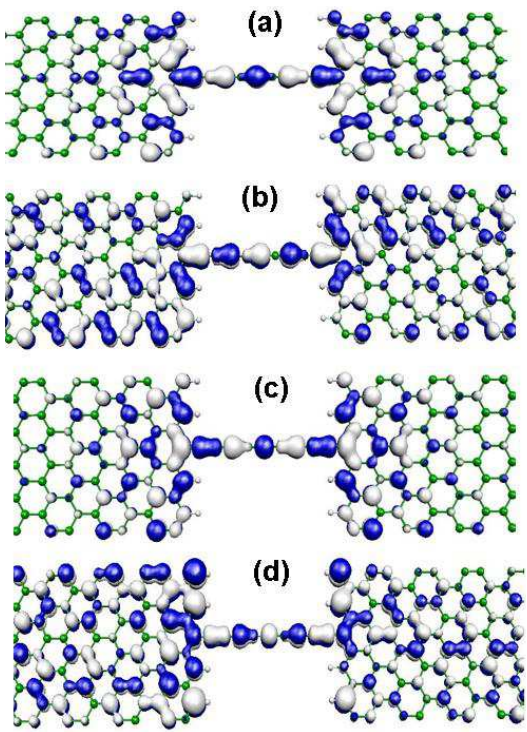}
\caption{(Color online) The real part of the minority spin
eigenstates of the C-9 system (a,c) and transmission eigenchannels
(b,d) at the resonances indicated by 1 and 2 in Fig. \ref{gmr},
respectively.}\label{sp_eig}
\end{center}
\end{figure}
Transmission occurs when graphene states are in resonance with the states of the atomic
chain. This is seen from the projected density of states (PDOS) of chain atom
$\pi_z$-orbitals plotted in Fig. \ref{gmr} (middle), which shows a clear correlation
with the transmission. The height of the peaks in transmission reflects the linear DOS
of graphene. The chain states differ from those of a corresponding free chain as the
chains in our case are found to strongly hybridize with the graphene edges. This is also
reflected in the magnetic moment of the contact edge atom being lower by roughly 0.2
$\mu_B$ compared to the edge atom furthest away from the chain. The
spin-splitting of the transmission curves originates from the spin-polarization of the
two zig-zag edges. The energetics favors a ferromagnetic chain-mediated coupling between the edges,
e.g., $E_G (AFM)-E_G(FM)=0.29$ eV for a nine-atom chain where $E_G (AFM)$ is the total energy of the AFM solution (See also S1).
A (weak) external magnetic field is needed to define a preferred spin orientation in an
experiment. In our calculations we consider only the ferromagnetic case. The magnetic moment of the chain, which is $N$ dependent,
almost entirely occurs in the $\pi_{x}$-orbitals and has little influence on the
transport.

The resonating states are investigated further in Fig. \ref{sp_eig},
using the minority spin states of the C-9 system as an example. In
(a) and (c) are shown the real part of the eigenstates of the total
system with energies corresponding to the peaks indicated in Fig.
\ref{gmr} by 1 and 2, respectively. The eigenstates  show a strong
hybridization between the edges and the chain. The transmission
eigenchannels at the resonances are shown in (b) and (d),
respectively. Except for a phase shift, the matching with the
eigenstates is clear. At energies corresponding to zero transmission
we find no eigenstates involving the $\pi_z$-orbitals. Calculations
on systems with wider leads confirm that hybridization only involves
the part of the edge close to the chain. The same analysis can be
carried out for the other resonances, and for both spins.


\begin{figure}[tbh]
\begin{center}
\includegraphics[width=0.8 \columnwidth,angle=0,clip]{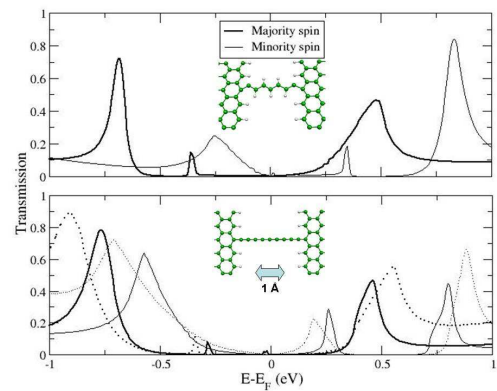}
\caption{(top) The transmission for a zig-zag shaped carbon chain
which is partly H terminated. (bottom) The transmission of the C-9
system stretched by 1 \AA. The results in the unstretched case are
shown for comparison (dotted line).} \label{mol}
\end{center}
\end{figure}
\begin{figure}[]
\begin{center}
\includegraphics[width=0.8 \columnwidth,angle=0,viewport= 0 0 700 500,clip]{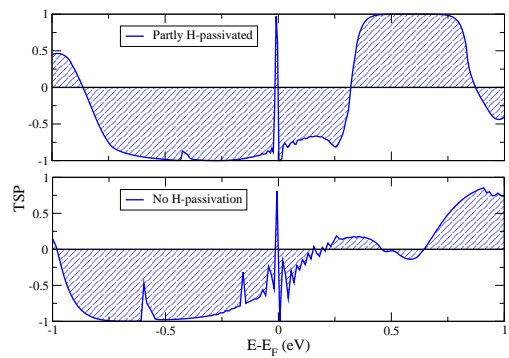}
\caption{TSP for a 9 atom chain connected to partly
H-passivated(top) and unpassivated (bottom) edges.} \label{h_test}
\end{center}
\end{figure}

We have tested the robustness of the spin-polarization effect in
various ways (details available in S1): (i) applying a bias voltage of 0.5 eV (ii) gating with a
lateral ($z$) field of 0.25 V/\AA~; (iii) adding a charge of -0.5 e;
and (iv) doubling the electrode width which reduces the relative
weight of the chains along the edge. Also, the electrodes in a real
sample are not semi-infinite sheets, but rather large ribbons or
flakes. We therefore also examined 9-atom wide fully H-passivated
semi-conducting armchair ribbon electrodes. In that case the details
in the transmission close to $E_F$ are influenced by the band gaps
in the leads, but the spin-polarization of the resonances is still
present. All these tests confirm the qualitative picture described
above.

The fact that the spin-polarization of the transmission is caused by an intrinsic
property of the zig-zag edge suggests that the high TSP values might be a generic
property of these systems. The spin-polarization effect should then occur also for all
carbon molecules which make an $sp^2$ bond to the zig-zag edge, and conduct via the
$\pi_z$-orbitals.
In order to test this hypothesis we have performed calculations for
a system where the linear chain is replaced with a partly
H-passivated zig-zag shaped chain as shown in the inset in Fig.
\ref{mol}, upper panel. Again, the splitting of the large resonances
appears in the same manner as for the linear chain, however even
more strongly in this case. Yet another possibility for tuning the
TSP is to mechanically change the electrode distances which induces
changes in the chain states. This is shown in Fig. \ref{mol}, lower
panel, where the C-9 system has been stretched by 1 \AA.

Finally, we comment on the influence of defects and edge passivation. Edge disorder has
been shown to affect the transport properties of graphene ribbons
\cite{Han2007,Gunlycke2007,Areshkin2007,Evaldsson2008,Xu2009}. However, the crucial
point in our case is whether the magnetisation is quenched or not. The chain acts as a
defect which is reflected in the mentioned significant lowering of the magnetic moment
of the contact atom. However, we find that the magnetisation along the - relatively
short - edge is rapidly restored away from the contact atom, in agreement with other
studies \cite{Jiang2007,Wimmer2008}. We thus infer that the effect is expected even in
the presence of nearby defects. The extent of edge passivation in actual samples is
uncertain \cite{Girit2009,Wassmann2008}. In the results presented so far, we have
considered full H passivation. In Fig. \ref{h_test} we show the TSP for the C-9 system
in the case where two H atoms has been removed from one edge so half of the edge atoms
are passivated, as well as the case of no passivation at all. The TSP is seen to be
largely intact below $E_F$, but reduced above $E_F$ compared to the case of full
passivation shown in Fig. \ref{gmr}. The TSP above $E_F$ can, however, be tuned as
explained above.
\par{\it Conclusion.}\label{sec_conc} We have presented {\it ab initio} results
for the electron transport properties of atomic carbon chains between zig-zag edges of
graphene electrodes. We find a complete spin-polarization of the electron transmission
in large energy ranges around the Fermi level owing to the intrinsic spin-polarization
of the zig-zag edges of graphene. Transmission occurs as graphene states are in
resonance with the strongly hybridized edge and chain $\pi_z$-orbital states. The
spin-polarization of the transmission can be controlled by manipulating the chemical
structure of the chain or a similar carbon molecule, applying mechanical strain, or
changing the position of the Fermi-level by doping or gating.

\par{\it Acknowledgments.} Computational resources were provided
by the Danish Center for Scientific Computations (DCSC). APJ is
grateful to the FiDiPro program of Academy of Finland.

\bibliography{./test}
\clearpage
\onecolumngrid

\begin{center}
\textit{\huge{Supplementary material}}
\end{center}
\begin{figure}[h!]
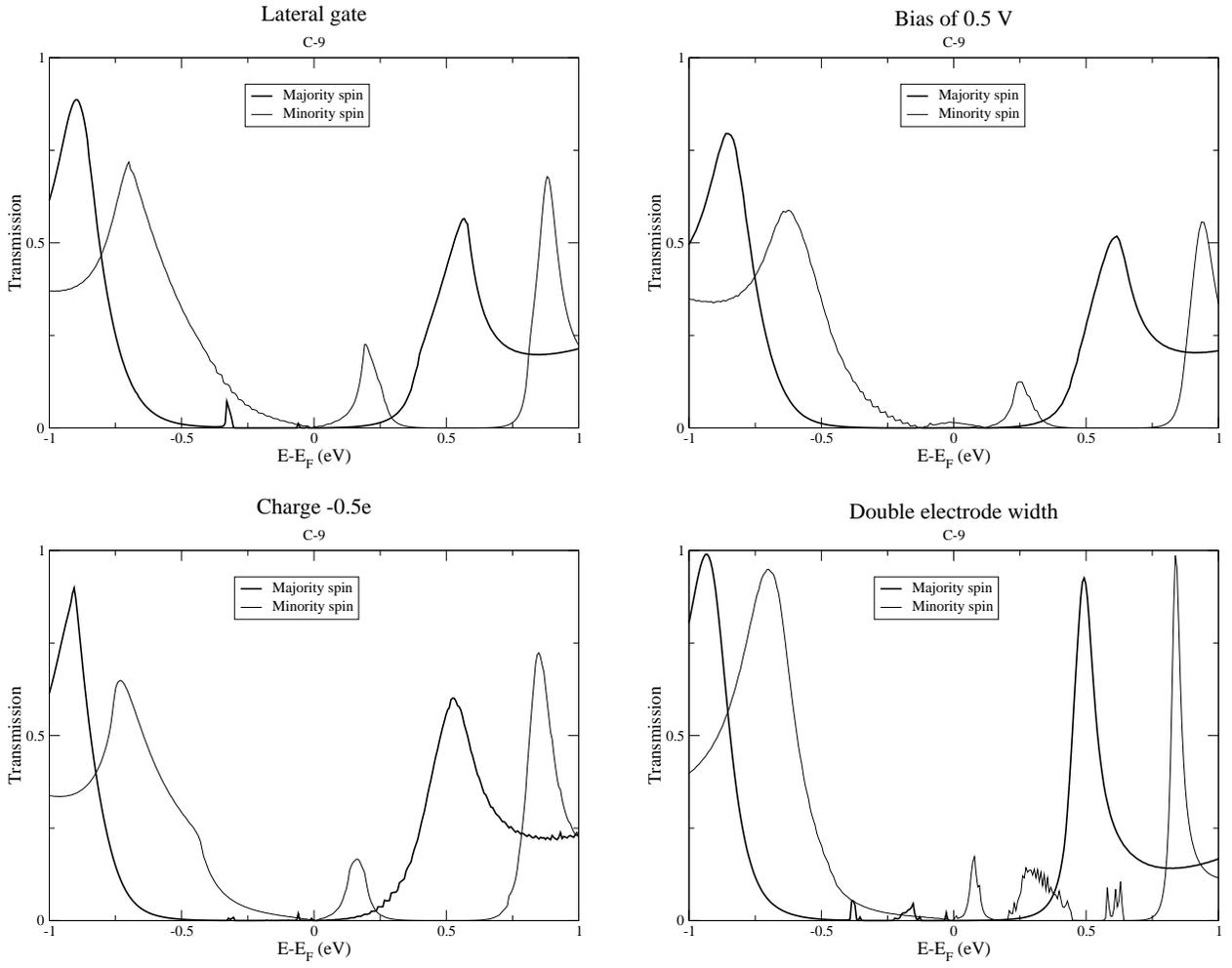

\begin{center}
\includegraphics[width=0.48 \columnwidth,angle=0,viewport=0 3 730 570,clip]{./gated_trans.eps}
\includegraphics[width=0.48 \columnwidth,angle=0,viewport=0 3 730 570,clip]{./bias.eps}
\includegraphics[width=0.48 \columnwidth,angle=0,viewport=0 3 730 570,clip]{./charge.eps}
\includegraphics[width=0.48 \columnwidth,angle=0,viewport=0 3 730 570,clip]{./dew.eps}
\caption{The electron transmission for the C-9 system when (top left) applying a bias voltage of 0.5 eV, (top right) gating with a lateral ($z$) field of 0.25 V/\AA~, (bottom left) adding a charge of -0.5e, (bottom right) doubling the electrode width which reduces
the relative weight of the chains along the edge.  }
\label{}
\end{center}
\end{figure}

\begin{figure}[h!]
\begin{center}
\mbox{
\subfigure[]{\includegraphics[width=0.44 \columnwidth,angle=0,viewport=0 3 560 510,clip]{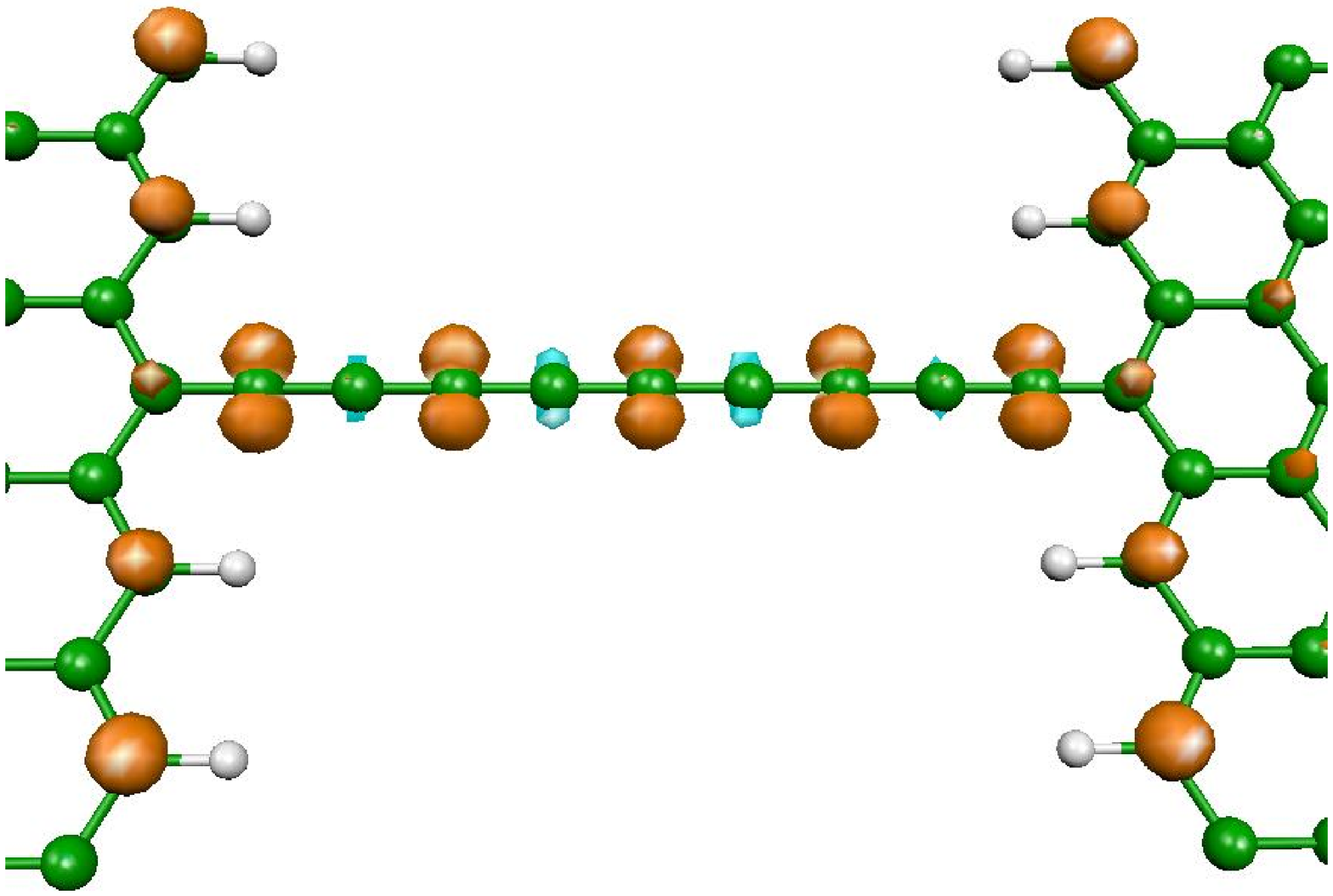}}\quad \quad \quad \quad
\subfigure[]{\includegraphics[width=0.44 \columnwidth,angle=0,viewport=0 3 560 510,clip]{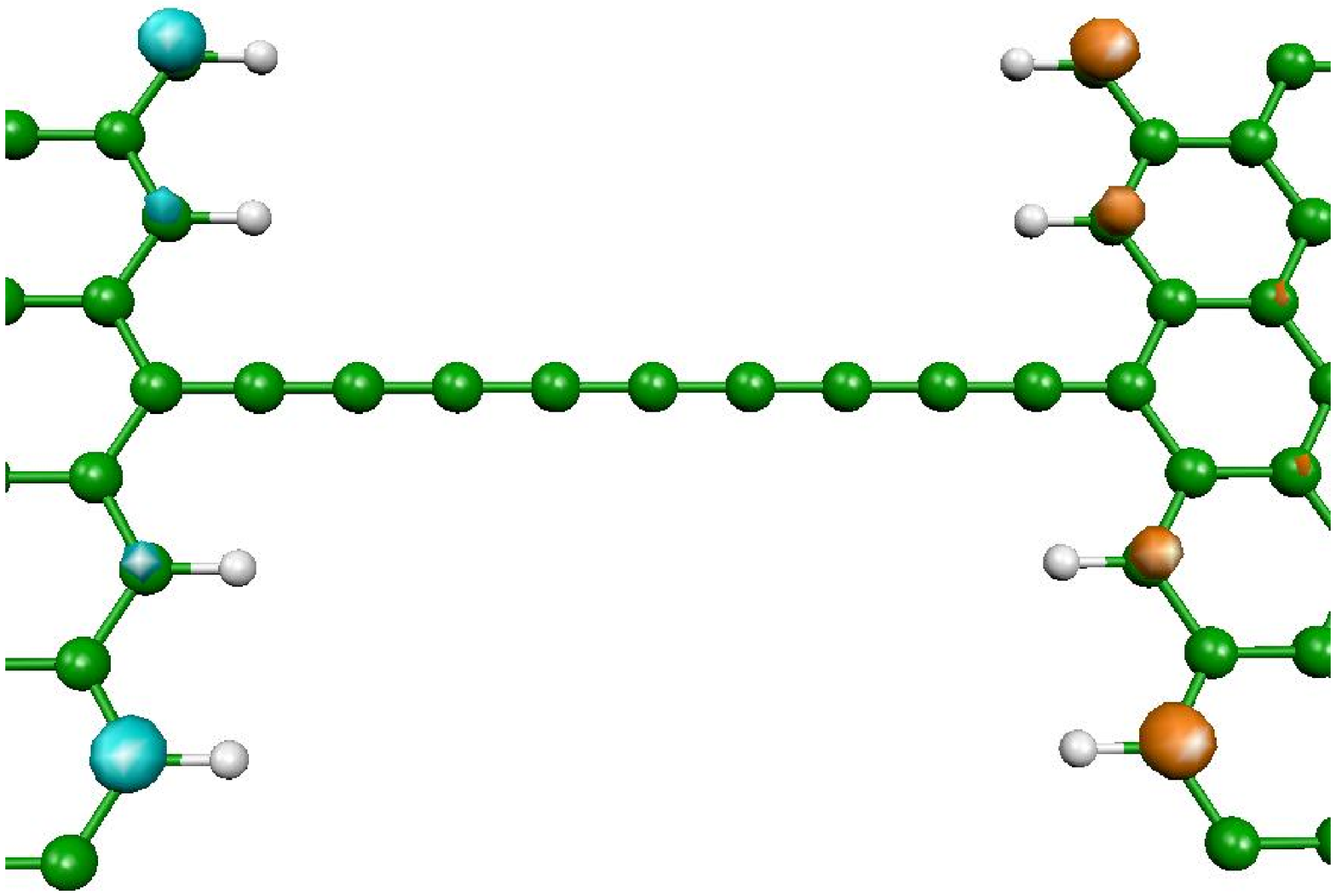}}
}
\caption{The difference in majority and minority spin densities for the C-9 system. Bronze (turquoise) indicates surplus of majority (minority) spin. (a): The FM solution between edges. The spin resides on the $\pi_z$ - orbitals at the edges and predominantly on the $\pi_x$ - orbitals in the chain. (b): An AFM solution obtained by initially breaking the FM symmetry by hand. Notice, that the chain is largely spin unpolarized. We obtain $E_G^{C-9} (AFM)-E_G^{C-9}(FM)=0.29$ eV and $E_G^{C-8} (AFM)-E_G^{C-8}(FM)=0.31$ eV. The connection of the two edges via the chain thus seem to favor a FM solution.}
\label{}
\end{center}
\end{figure}
\begin{figure}[tbh]
\begin{center}
\includegraphics[width=0.6\columnwidth,angle=0,viewport= 0 0 670 350, clip]{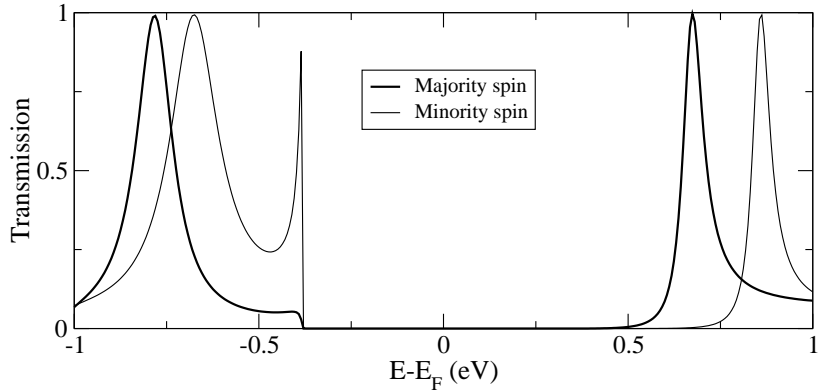}
\caption{The transmission for a 9 atom chain connected to fully H-passivated armchair semiconducting ribbon electrodes with a width of 9 rows of atoms.}
\label{ribbon}
\end{center}
\end{figure}

\begin{figure}[tbh]
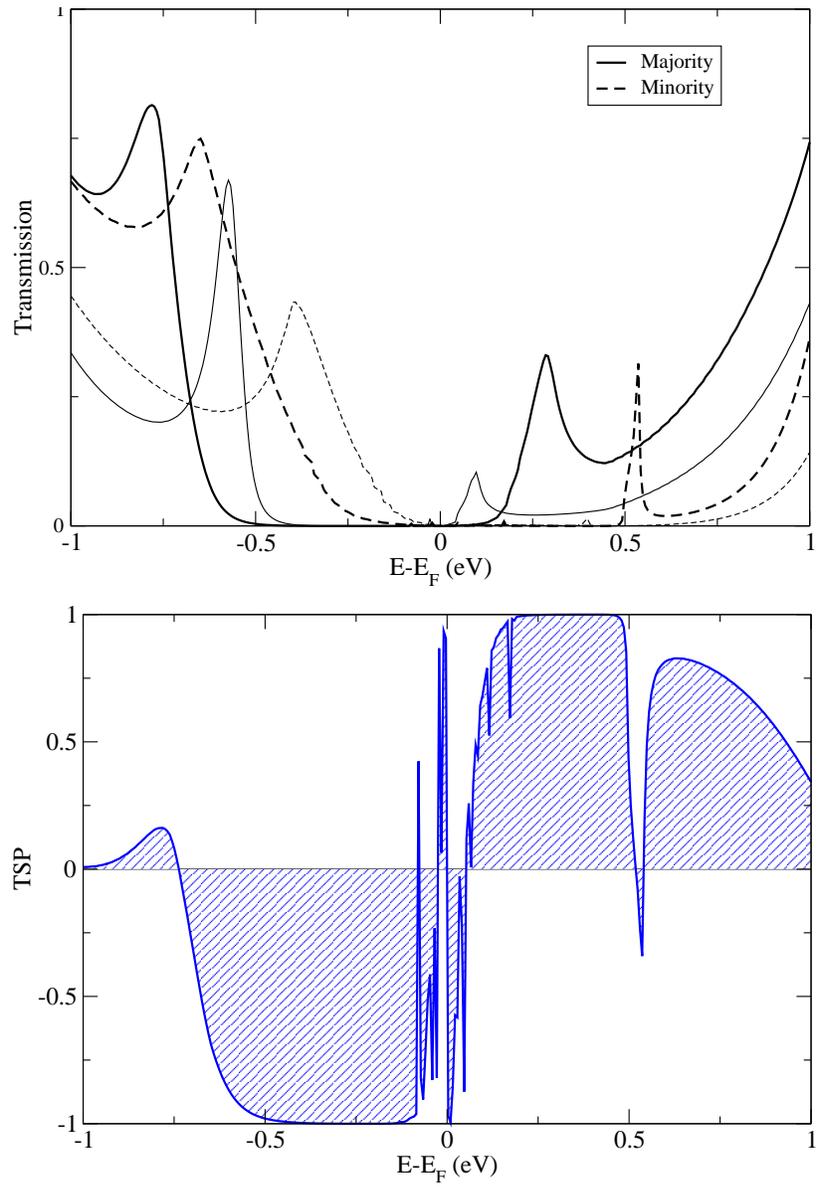

\begin{center}
\includegraphics[width=0.6\columnwidth,angle=0,viewport= 0 0 670 480, clip]{./compare.eps}
\includegraphics[width=0.6\columnwidth,angle=0,viewport= 0 0 680 500, clip]{./dimer_tsp.eps}
\caption{The transmission (top) and TSP (bottom) for the C-8 system where a dimerisation has been inserted by hand with single/triple-bond lengths of 1.26/1.36 \AA.}
\label{dimer}
\end{center}
\end{figure}

\end{document}